\documentclass[prd, twocolumn, nofootinbib, floatfix]{revtex4-1}

\usepackage[mathscr]{euscript}
\usepackage{amsmath}
\usepackage{graphicx}
\usepackage{dcolumn}
\usepackage{bm}
\usepackage{epsfig}
\usepackage{amssymb,latexsym,mathrsfs}
\usepackage{graphicx}
\usepackage{color}
\usepackage{hyperref}
\usepackage{float}
\usepackage{diagbox}

\hypersetup{
    colorlinks=true,
    linkcolor=red,
    citecolor=blue,
} 

\newcommand{\be}{\begin{equation}}
\newcommand{\ee}{\end{equation}}
\newcommand{\bs}{\begin{split}} 
\newcommand{\bea}{\begin{eqnarray}}
\newcommand{\eea}{\end{eqnarray}}

\renewcommand{\d}[1]{\ensuremath{\operatorname{d}\!{#1}}}

\begin{document}

\title{Schwarzschild Metric with Planck Length} 
\author{Michael R.R. Good${}^{1,2}$}
\author{Eric V.\ Linder${}^{2,3}$} 
\affiliation{${}^1$Physics Department, Nazarbayev University, Nur-Sultan, Kazakhstan\\
${}^2$Energetic Cosmos Laboratory, Nazarbayev University, Nur-Sultan, Kazakhstan\\ 
${}^3$Berkeley Center for Cosmological Physics \& Berkeley Lab, University of California, Berkeley, CA, USA
}

\begin{abstract} 
We present a modified Schwarzschild solution for a model of evaporation of a black hole with information preservation. By drawing a 
direct analogy to the quantum pure 
accelerating mirror (dynamical Casimir effect of a 1D horizon), we derive a 
Schwarzschild metric with not only the 
usual Schwarzschild radius but an additional 
length scale related to the Planck length. 
The black hole has thermal particle production that leads to complete evaporation of the black hole, resulting in non-divergent entanglement entropy, Page curve turn-over, and an asymptotic quantum pure state with no information loss. 
\end{abstract} 

\date{\today} 

\maketitle

\section{Introduction} 

Particle production from black holes, or more generally curved 
spacetime, is a fascinating process, bringing together quantum 
physics and gravity (or its Equivalence Principle partners of 
acceleration or curvature). 
Particles can be produced through 
\begin{itemize}
\item Expanding Cosmologies (Parker effect \cite{Parker:1968mv})
\item Black Holes (Hawking effect 
\cite{Hawking}) 
\item Uniform Acceleration Radiation (Unruh effect 
\cite{Unruh}) 
\item Moving Mirrors (Davies-Fulling effect 
\cite{DeWitt,Davies:1976hi,Davies:1977yv}) 
\end{itemize} 

Beyond the mere existence of this 
interesting physical process, we might seek for two characteristics 
with long ties to the development of quantum theory: finite 
global particle count (e.g.\ avoidance of divergent numbers of 
soft photons \cite{soft}) and a thermal (blackbody) radiation spectrum. 
Furthermore, the information carried by the radiation, and whether it is preserved or lost, is the subject of intense scrutiny \cite{Chen}. 

Here we look at the simplest black hole case, that of the Schwarzschild metric, and explore a model for its formation and evaporation with information preservation. We motivate that this leads to introduction of a second length scale beyond the horizon size (which suggestively can be related to the Planck length), and explore what happens near the horizon and asymptotically, as well as when the horizon size approaches the new ``quantum'' length scale after long evaporation. 

In Section~\ref{sec:schpl} we introduce the model and compute the spacetime geometric quantities. We study the particle production in Sec.~\ref{sec:particle} in various limits, including the spectrum, total energy, and entropy, and quantum pure states. 
Section~\ref{sec:concl} summarizes and concludes.

\section{Schwarzschild and Planck} \label{sec:schpl} 

The Schwarzschild metric describes a static black hole spacetime, and is 
central to exploring the physics of black holes, one of the most challenging 
subjects in physics. It can be derived based on certain symmetry principles of 
the spacetime, and Einstein's equations for gravity. It is a key stage on which 
we explore extremes of gravitational physics, and its frontier with quantum 
physics. A black hole produces particles \cite{Hawking} from the spacetime by 
quantum physics, but neither the metric nor Einstein's equations reflect 
quantum physics. The particles radiate \cite{Unruh, DeWitt} to infinity in the external spacetime, 
yet the external state is taken to be vacuum. What happens when the black 
hole radiates all its energy is unknown, and the subject of information paradoxes \cite{Chen,Hayden:2007cs}. 
Here we attempt to explore methods \cite{Parker_book,Fabbri} for probing these inconsistencies and puzzles.

\subsection{Schwarzschild} 

In the classical picture, the Schwarzschild metric can be written as 
\be 
ds^2 = -f_s dt^2 + f_s^{-1}dr^2 + r^2 d\Omega\,,\label{CS} 
\ee 
where the angular part of the metric is spherically symmetric, $d\Omega \equiv d\theta^2 + \sin^2\theta d\phi^2$, and 
\be 
f_s=1-\frac{r_s}{r}\,, 
\ee 
where $r_s=2M$ is the Schwarzschild radius (related to the black hole mass $M$). 

For studying the structure of spacetime, it is convenient to follow null rays 
and remove the coordinate singularity at $r=r_s$. This can be done with 
the Regge-Wheeler tortoise coordinate 
\be 
r^* = r+ r_s \ln \left|\frac{\Delta}{r_s}\right|\label{tort}\,, 
\ee 
where $\Delta\equiv r-r_s$. Rays of constant $t\pm r^*$ correspond to 
ingoing and outgoing null geodesics, and we can define 
\be 
v = t+r^*\,, \quad u = t-r^*\,,
\ee 
for outside coordinates, and the equivalent $V$ and $U$ for inside coordinates, $T\pm r$. 
The derivative of the transformation gives the metric, 
\be 
\frac{dr^*}{dr} \equiv f_s^{-1}\,. 
\ee 

The canonical collapse prescription (see e.g.\  \cite{purity, Fabbri}) for a shell of matter at $v_0$, where the horizon $v_H=v_0-2r_s$, gives the matching condition utilizing the regularity condition effect on the form of the modes, $v\leftrightarrow U$ \cite{Fabbri} 
\be 
u(U) = U - 4M \ln\left|\frac{U}{4M}\right| 
= U -\frac{1}{\kappa}\ln\left|\kappa U\right|\,, \label{eq:classu} 
\ee
where $\kappa=1/(4M)=1/(2r_s)$ is the surface gravity of the black hole. 
Note that $du/dU = f_s^{-1}$ and we are free to set $v_H=0$ without 
loss of generality. 

At this point, we can draw attention to the analogy of a black hole, and its 
quantum physical particle production, to the Davies-Fulling effect \cite{Davies:1976hi,Davies:1977yv} for particle 
production by an accelerating mirror. This basically arises 
due to boundary conditions imposed on the spacetime, a dynamical Casimir effect, 
in much the same way that the black hole horizon is a boundary. Crafting the mirror trajectory to match the null coordinate form $u(U)$ in Eq.~(\ref{eq:classu}), 
i.e.\ identifying this as 
the trajectory of the origin in null coordinates of the mirror \cite{Good:2016MRB}, 
\be 
\tau(v) = v - \frac{1}{\kappa} \ln |\kappa v|\,,\label{f(v)BHM}\ee 
gives what is known as the black mirror \cite{MG14one,MG14two,Good:2016LECOSPA}, which has the same late-time eternal thermal 
particle production as a black hole. 

Let us review our path. From the metric derived from the symmetries of spacetime 
and gravity, we moved to tortoise coordinates, matching conditions for collapse, 
and an analogy to particle production from a moving mirror. In the next 
subsection  we will reverse this journey in the hope of learning something 
fundamental about spacetime around black holes with quantum physics included.

\subsection{Planck} 

The problem with the black mirror is that the analogy is too perfect. The 
particle production at late times is in eternal equilibrium, giving rise to an infinite 
number of particles and infinite total energy, as well as a black hole that simply radiates forever. The information paradox is a concern here (for much of the community at least, e.g.\ harvesting entanglement via the black mirror \cite{harvest}) because violation of unitarity may contradict basic principles of probability and quantum theory. This black mirror-black hole model does not sufficiently clarify the connection between quantum physics and gravity. 

However, recently a different sort of moving mirror trajectory was discovered 
\cite{GLW}, the quantum pure black mirror. This has both finite number of 
particles produced and finite total energy, acts like a black hole that 
evaporates completely \cite{walkerdavies} without leaving a remnant, and produces quasi-thermal radiation that 
ends in a pure quantum state. Entropy \cite{geoentropy,paper3,yeomentropy,movingmirrorentropy} is well behaved and there is no information 
loss. 

The trajectory of the quantum pure black mirror is 
\be 
\tau(v) = v- \frac{1}{\kappa}\sinh^{-1}|g v|\,, \label{f(v)FTP} 
\ee 
where there is now a new parameter $g$ in addition to $\kappa$. Basically 
$\kappa$ will determine the amplitude of the particle flux produced (in the 
same way that in the black hole case the surface gravity or mass does), while $g$ will 
factor into the evaporation of the black hole, i.e.\ the lifetime. Just as 
we write $\kappa=1/(2r_s)$, we can write $g=1/(2l)$ and anticipate that $l$ 
will be related to the Planck length $l_P=\sqrt{\hbar G/c^3}$. This will allow 
quantum physics and gravity to both enter the picture. 

Now we reverse our journey of the last subsection. We take the collapse 
condition in the black hole case in exact analogy to Eq.~(\ref{f(v)FTP}), so 
\be 
u(U) = U- 2r_s \sinh^{-1}\left|\frac{U}{2l}\right|\,, \label{matching} 
\ee 
and define from this a quantum tortoise coordinate 
\be 
\bar{r}^* = r+ r_s \sinh^{-1}\left|\frac{\Delta}{l}\right|\,.  
\label{newtort} 
\ee 
Up to a constant\footnote{The constant can be found by setting $\bar{r}^* = r^* = 0$  and $r =1+W(1/e)$ in units of $r_s$. Here $W$ is the Lambert $W$ function.}, 
the quantum tortoise coordinate, Eq.~(\ref{newtort}), goes to the classical tortoise coordinate, Eq.~(\ref{tort}), 
\be 
\lim_{\mathscr{R}\rightarrow 0} \bar{r}^* = r^*, \quad \textrm{for} \quad r > r_s\,,  
\ee
where $\mathscr{R} \equiv l/r_s$. See the next subsection 
for more discussion. 

Using that $d\bar{r}^*/dr \equiv \bar{f}^{-1}$ we now have our quantum 
replacement for the Schwarzschild metric: 
\be 
ds^2 = -\bar{f} dt^2 + \bar{f}^{-1}dr^2 + r^2 d\Omega\,, \label{QS} 
\ee 
where instead of the usual $f = 1 - r_s/r$, we have 
\be 
\bar{f} \equiv 1 - \frac{r_s}{r_s+ \sqrt{\Delta^2+l^2}}\,, \label{fbar} 
\ee 
with $\Delta \equiv r-r_s$.  Clearly, when $l \rightarrow 0$, then $\bar{f} \rightarrow f$ gives the classical Schwarzschild solution. However, now there 
are further effects expected when $r-r_s\lesssim l$, e.g.\ within some Planck 
lengths of the horizon. And if $l$ is indeed connected with the Planck length, 
then this is a quantum metric in the sense that $g_{\mu\nu}$ includes $\hbar$. 

To explore the effects of this new metric further, we investigate the 
Kretschmann scalar, which gives an invariant description of the curvature 
and singularities, the Ricci scalar, and the Einstein tensor. To leading 
order in  $l/r_s$ beyond the classical values (see Appendix~\ref{sec:apxcnx} 
for full expressions), near $r=r_s$, the scalars are,
\bea K &=& \frac{12 r_s^2}{r^6} -\frac{4 l^2 }{r_s^3 \Delta ^3}+ \dots,\\ 
R &=& 0 - \frac{l^2}{r_s\Delta{}^3}  +\dots.
\eea
while the Einstein tensor is,
\bea
G_{\mu\nu} &=& 0 +\\ 
&\quad& \left(
\begin{array}{cccc}
 \frac{\Delta ^2}{r_s^2} &  &  &  \\
  & -1 &  &  \\
  &  & r_s^2 &  \\
  &  &  & r_s^2\sin ^2\theta \\
\end{array}
\right)\frac{l^2}{2r_s\Delta^3}+\dots. 
\eea 

Indeed when $l$ vanishes we recover the classical Schwarzschild vacuum 
solution. However, there are several new aspects that arise with finite $l$. 
This is no longer a vacuum solution, i.e.\ $G_{\mu\nu}\ne0$. That makes sense: 
for one thing the classical Schwarzschild metric is the unique vacuum solution 
for this spacetime, and for another we know there is particle production -- 
we must break the vacuum solution. In a sense, since black holes 
radiate then the classical Schwarzschild metric should not be the true metric. We also see that when we are within some Planck lengths of the horizon, 
i.e.\ $\Delta=|r-r_s|\lesssim l$, that the Kretschmann and Ricci scalars 
deviate strongly from the classical behavior. We discuss this in detail in Appendix~\ref{sec:apxcnx}.

\subsection{Classical and Quantum, Black Hole and Mirror} 

\begin{figure}[ht]
\centering 
\includegraphics[width=3.2in]{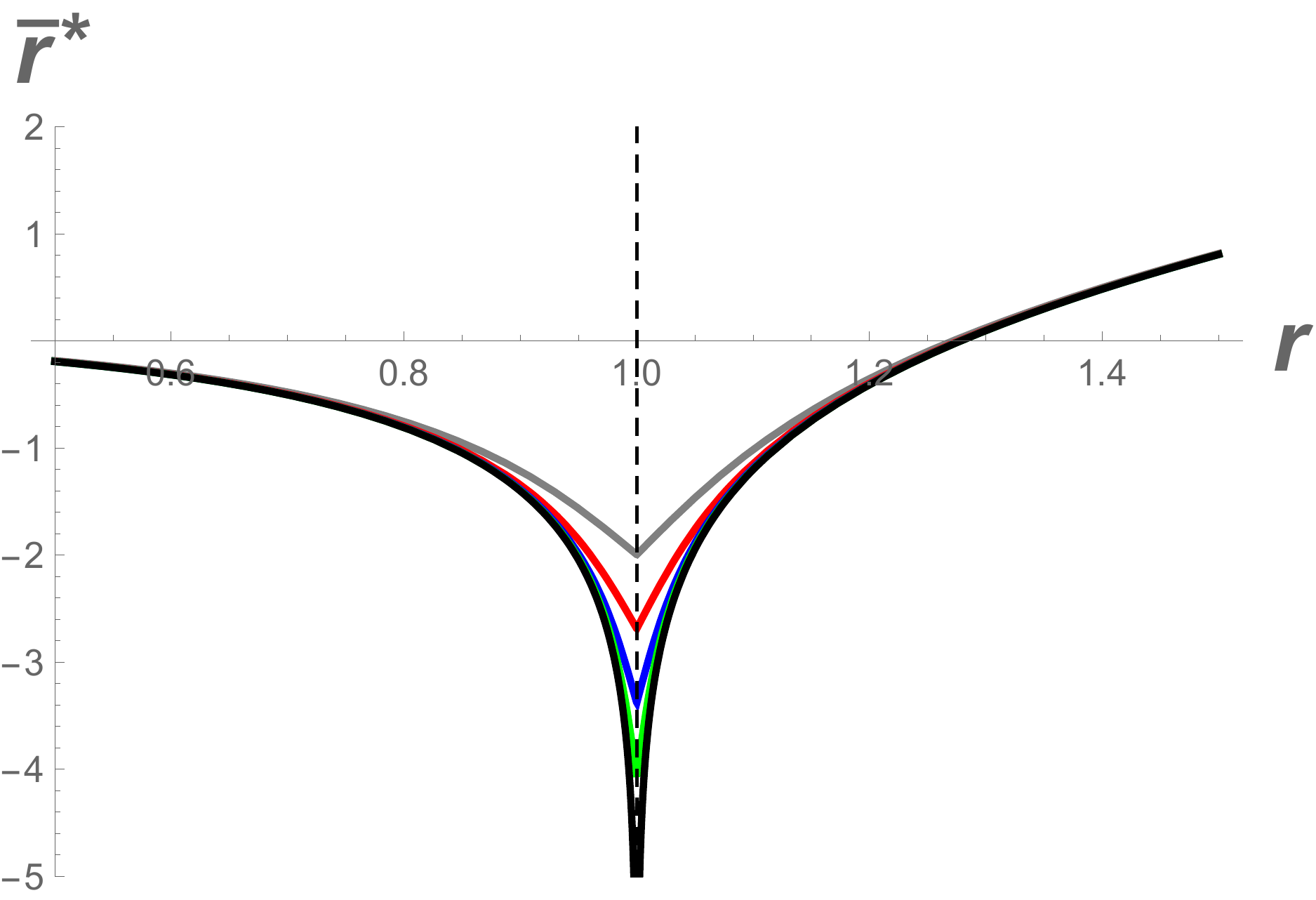} 
\caption{The classical and quantum tortoise coordinates  are 
plotted vs standard coordinate $r$, measured in units of $r_s$. 
The black line is the usual classical Regge-Wheeler tortoise 
coordinate, which diverges at $r=r_s$. The new quantum tortoise 
coordinate, Eq.~(\ref{newtort}) (up to a constant), stays 
finite and involves a quantum scale $l$. Gray, red, blue, and 
green curves show the quantum tortoise coordinate for values 
of $l$ starting at $l=0.1$ (in units of $r_s$) and halved 
sequentially, respectively (so the green curve is for $l=0.0125$). 
}
\label{fig:Tortoise_Plot} 
\end{figure}

Our attempt to move from the classical to the quantum used the 
moving mirror analogy to suggest a new, quantum tortoise 
coordinate. Both the classical and quantum tortoise coordinates 
are plotted in Figure ~\ref{fig:Tortoise_Plot}.  
Note the quantum tortoise coordinate removes the 
divergence of the classical tortoise coordinate.

\begin{figure}[ht]
\centering 
\includegraphics[width=2.3in,height=2.3in,keepaspectratio]{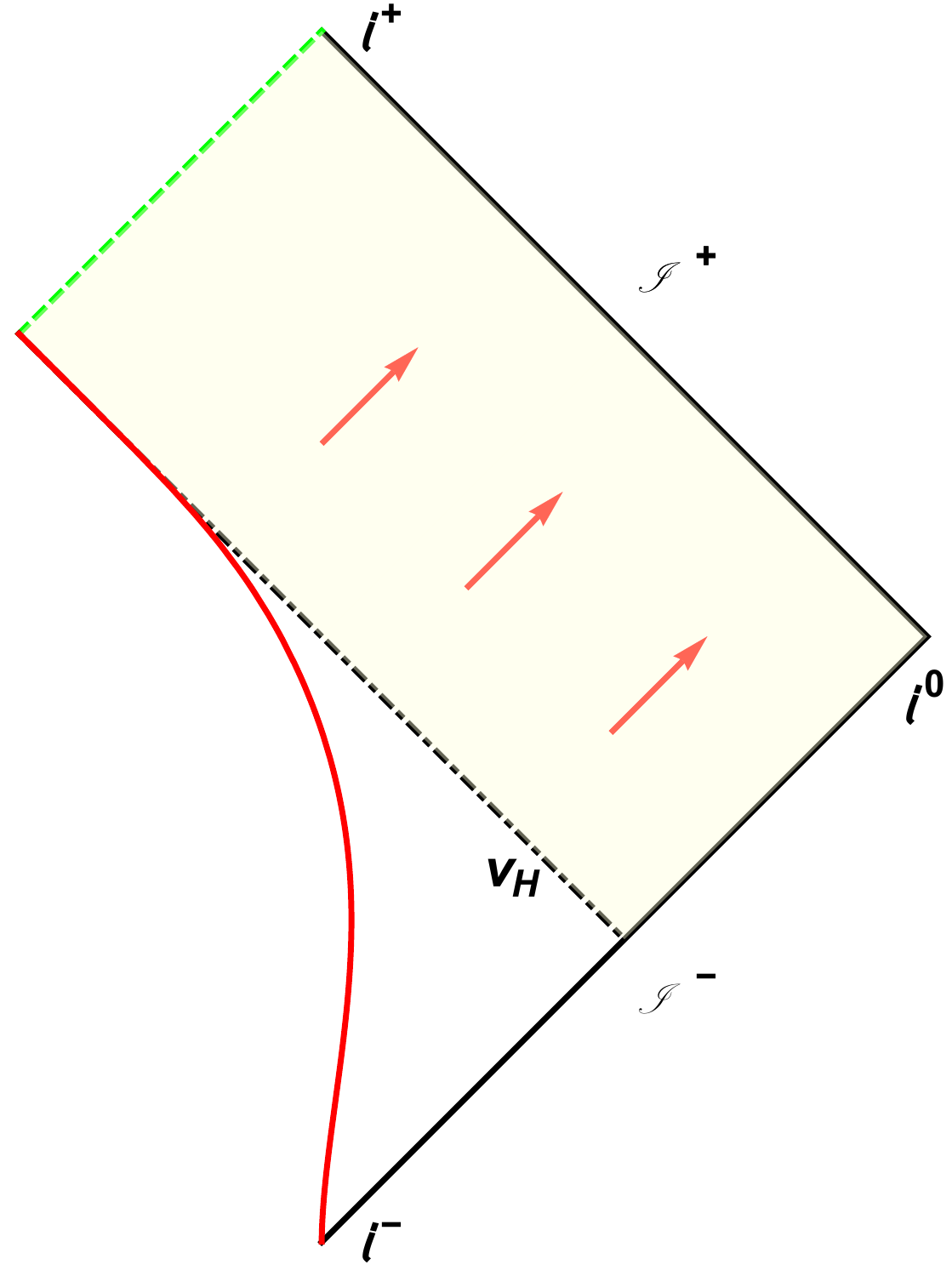} \;\;\;
\includegraphics[width=2.3in,height=2.3in,keepaspectratio]{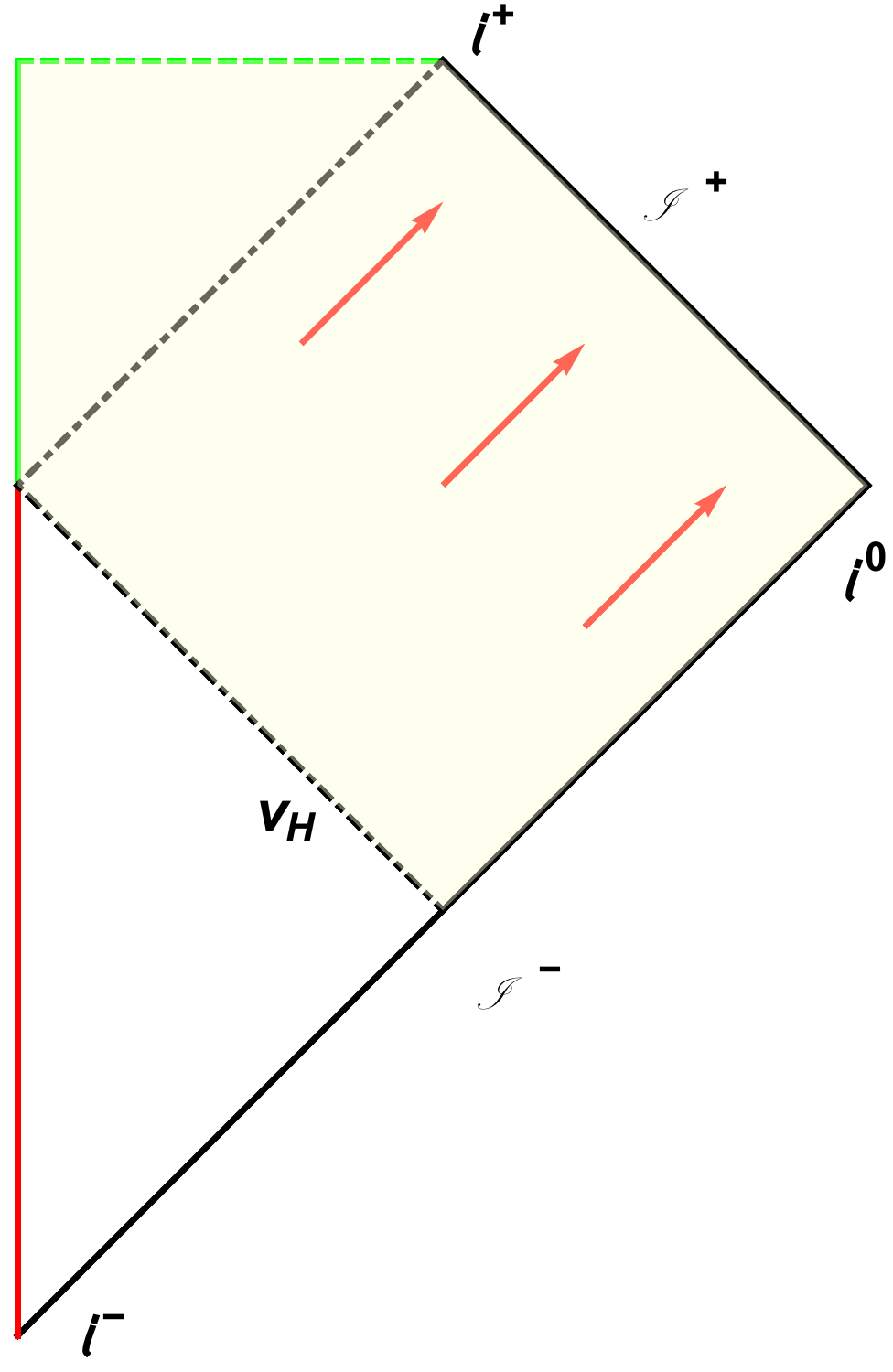} 
\caption{\textbf{Left}: The 1+1 dimensional Penrose (conformal) diagram for the black hole collapse analog mirror, `black mirror' \cite{MG14one,MG14two,Good:2016LECOSPA,Good:2016MRB}, or BHC trajectory \cite{harvest}.  The red line is the accelerated boundary, computed with $\kappa=4$ for illustration. The dot-dashed black line is the advanced time horizon, $v_H= 0$.  The green dashed line is the upper-half of the left future null-infinity surface, $\mathscr{I}_L^+$, which has been left unlabeled.  It is the singularity analog. Notice the light yellow shaded region is where left-moving modes never reflect off the mirror and fall inside, never to return again, i.e. they do not become right-movers.  The red arrows represent radiation emitted by the mirror, or, if preferred, they can be left-mover modes that reflected off the mirror into right-mover modes reaching our observer at $\mathscr{I}^+$. \textbf{Right}: The usual 3+1 dimensional Schwarzschild causal structure for a black hole, captured in a Penrose diagram.  The green dashed line is the space-like singularity while the green solid line is $r=0$ where modes pass through (`reflect') but still hit the singularity.  The interesting aspect here is that there is no mirror counterpart for the solid green line.  The black dot-dashed line is the event horizon as usual, while the red arrows are again Hawking radiation.  The red arrows can also be seen as left-movers that passed through $r=0$, `regularity-reflected' and become right-movers.  They will eventually reach the observer at $\mathscr{I}^+$. The light yellow shaded region marks where left-movers eventually fall into the singularity.  
}
\label{fig:Black_Mirror} 
\end{figure}

The moving mirror analog to a black hole is useful not only in 
describing particle flux, but spacetime structure, 
singularities, and asymptotic 
conditions as well.  Figure~\ref{fig:Black_Mirror} 
illustrates the classical correspondence in terms of a 
conformal spacetime diagram, with the mirror on the left and 
the black hole on the right. Note the strict $v$-horizon at 
$v_H$ (black dotted line), which gives a divergent $\tau(v)$, signaling `incomplete' evaporation -- information loss.

\begin{figure}[ht]
\centering 
\includegraphics[width=2.3in,height=2.3in,keepaspectratio]{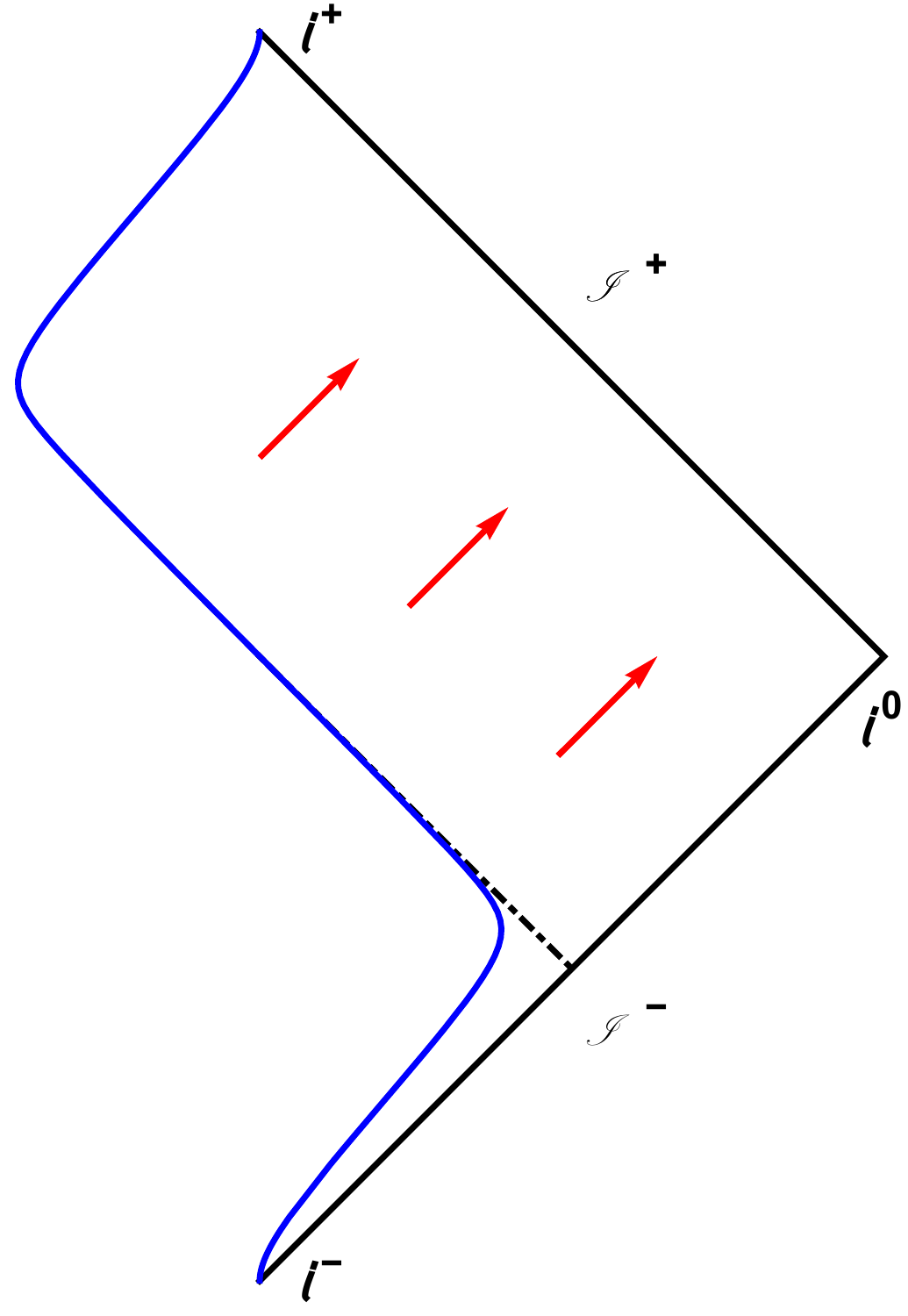}  \;\;\; \;\;\;
\includegraphics[width=2.3in,height=2.3in,keepaspectratio]{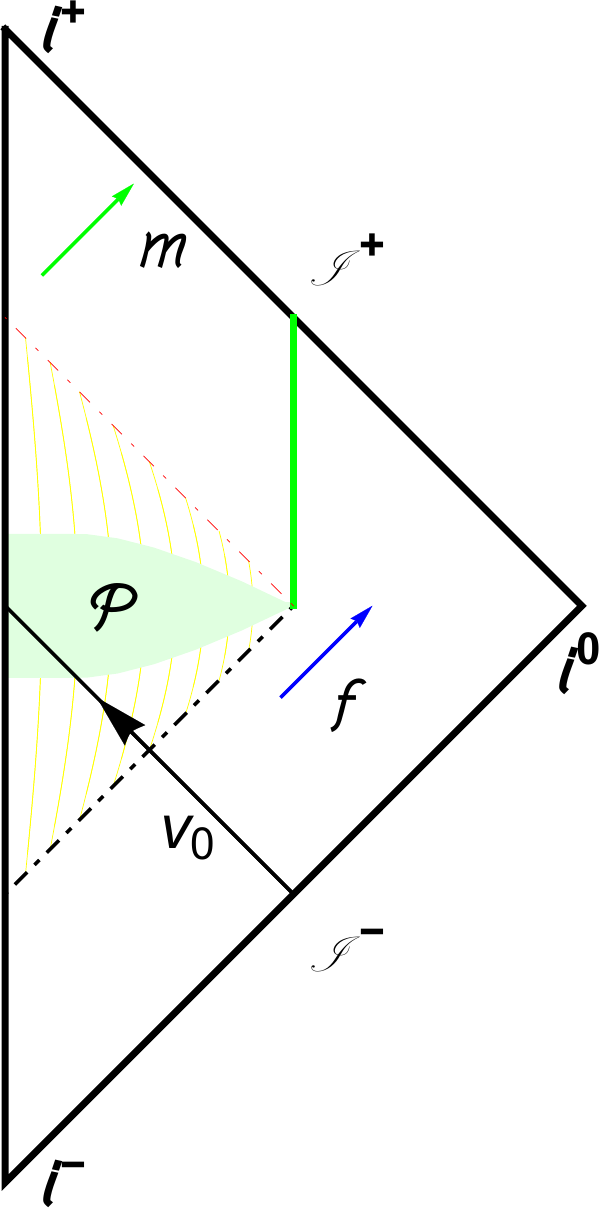} 
\caption{
\textbf{Left}: The 1+1 dimensional Penrose diagram for the quantum pure black mirror Eq.~(\ref{f(v)FTP}), blue line, computed with $\kappa=1$ and $g= 10^2$. The dot-dashed black line is the horizon, $v_H= 0$. All left-moving modes reflect off the mirror and become right-movers.  The red arrows are Hawking radiation emitted by the mirror or they can be left-movers that reflected off the mirror into right-movers reaching $\mathscr{I}^+$. \textbf{Right}: Penrose diagram for the no-remnant analog 3+1 dimensional black hole case with complete evaporation. Notice the non-redshifted outgoing modes labeled as $\mathit{m}$ with a green arrow, escaping to null-infinity $\mathscr{I}^+$, with the same plane wave form as ultra-early time modes, giving no hint that a black hole had ever lived.  In this case Hawking evaporation stops, all field modes passing through the thick green line after the null ray (dotted red line) undergo zero late-time redshift.  The blue arrow $f$ represents the red-shifted Hawking radiation flux.  The light green area labeled $\mathcal{P}$ is the Planckian region where quantum geometric effects are important. $\mathcal{P}$ has a non-zero, maximum, finite height centered at $r=0$.  All field modes can `reflect' off of $r=0$ and pass through the $\mathcal{P}$ region without loss of information, as they return to $\mathscr{I}^+$.  The yellow bands mark spacetime inside the black hole horizon.   }
\label{fig:FTP_Mirror} 
\end{figure}

Figure~\ref{fig:FTP_Mirror} by contrast shows the new 
``quantum'' situation. The lack of a strict $v$-horizon in Eq.~(\ref{f(v)FTP}), contrasted with Eq.~(\ref{f(v)BHM}), signals complete evaporation -- no information loss.  Moreover, there is no remnant (see the contrasting cases where the modes are affected long after the radiation stops, e.g.\  \cite{horizonless,GTC,universe,MG15}), since the field modes have the same early-time and late-time form (the mirror comes back to rest so 
there is no eternal redshift, e.g.\ \cite{purity, paper1,walkerdavies}).

\section{Particle Production} \label{sec:particle} 

The modes of the quantum field determine the particle 
production \cite{B&D}. One of the powerful aspects of the quantum pure 
black mirror is the analytic expression for the  
Bogolyubov beta coefficients \cite{GLW}, allowing simple 
computation of the particle production properties (e.g.\ the spectrum \cite{MPA}). The 
energy radiated also takes a simple expression (in the $1+1$ dimensional context, at least). In the following subsections we discuss several of these properties.

\subsection{Spatial and Temporal Limits} 

The field must be zero at the origin, as no field exists in $r<0$. Regularity forces the form of the field modes
\be \phi_\omega^{\textrm{out}} = \frac{1}{4\pi r \sqrt{\omega}}\left(e^{-i\omega u(v)}-e^{-i\omega u(U)}\right)\,,\ee
where $\omega$ is the frequency, to assume 
\be u(v) = u(U)\,.\ee
Therefore we write the matching condition Eq.~(\ref{matching}), as
\be u(v) = v - 2r_s \sinh^{-1}\left|\frac{v}{2l}\right|\,.\ee
This will be used in the modes to determine the behavior of the quantum field.   A complete cover of the collapse spacetime is given in Appendix \ref{cover}.

Armed with the matching condition, it is instructive to analyze the behavior of the modes at very early and at ultra-late times. We also look at a third temporal limit -- 
the equilibrium times.

\subsubsection{Early Times}
 At early times, when $v \rightarrow -\infty$, we have $u(v) \approx v$. This is analogous to the mirror in the asymptotic past being at rest. When $U\rightarrow -\infty$, the matching condition Eq.~(\ref{matching}) implies $u(U) \approx U$. Therefore at $\mathscr{I}^-$, early times, 
\be \phi^{\textrm{out}}_\omega \approx \frac{1}{4\pi r \sqrt{\omega}}\ e^{-i \omega v}, \ee
where the mode at $u\rightarrow -\infty$ is a pure positive frequency mode with respect to inertial time at $\mathscr{I}^-$. The Bogolyubov beta coefficients vanish and asymptotically early times exhibit no particle creation. There is nothing new here from the canonical case. There are no particles yet because the black hole has not formed, and its progenitor is only just beginning to collapse.

\subsubsection{Equilibrium Times}

With the regularity condition, the exact mode form assumes 
\be \phi_\omega^{\textrm{out}} = \frac{1}{4\pi r \sqrt{\omega}}\left(e^{-i\omega\left( v - 2r_s \sinh^{-1}\left|\frac{v}{2l}\right|\right)}-e^{-i\omega u}\right)\,,\ee
and at $\mathscr{I}^-$ where the $u$ piece has infinite oscillation, the mode
\be \phi_\omega^{\textrm{out}} = \frac{1}{4\pi r \sqrt{\omega}}\ e^{-i\omega\left( v - 2r_s \sinh^{-1}\left|\frac{v}{2l}\right|\right)}\,,\ee
displays a strong blueshift
in the exponent. Near the `horizon' $v_H = 0$, the mode becomes
\be \phi_\omega^{\textrm{out}} = \frac{1}{4\pi r \sqrt{\omega}}\ e^{-i\omega\left( 1+ \frac{r_s}{l}\right)v},\ee
typical of the process of black hole evaporation, which will largely modify the physics as calculated from the modes.  Since $r_s \gg l$, it is particularly easy to see the geometric modification of the field. The form of the modes at intermediate times gives non-zero beta coefficients, resulting in particle creation.  Moreover, the particular betas result in a quasi-thermal particle spectrum 
(see Sec.~\ref{sec:spec}).

\subsubsection{Ultra-Late Times} 

At ultra-late times, when $v \rightarrow +\infty$, we have $u(v) \approx v$, again, like the situation at early times. This is seen because $U\rightarrow +\infty$, the matching condition Eq.~(\ref{matching}), implies $u(U) \approx U$. Therefore at $\mathscr{I}^-$, ultra-late times, 
\be \phi^{\textrm{out}}_\omega \approx \frac{1}{4\pi r \sqrt{\omega}}\ e^{-i \omega v}, \ee
where the mode at $u\rightarrow +\infty$ is a pure positive frequency mode with respect to inertial time at $\mathscr{I}^-$. The beta coefficients vanish. At asymptotically ultra-late times there is no more particle creation, and one observes complete evaporation.  Unitarity is maintained.

\subsection{Radiation Stress Energy} 

Neglecting the time-independent Boulware vacuum polarization terms, specializing to conformal symmetry of $1+1$ dimensions, the time-dependent non-zero component of the normal-ordered (denoted by colons) stress tensor in the $in$-vacuum state is computed as the Schwarzian derivative \cite{Fabbri},
\be F(u) \equiv \langle in | :T_{uu}(u): |in\rangle = -\frac{\hbar}{24\pi} \{U(u),u\}.\ee
This is most easily done by utilizing the invariant,
\be \{u(U),U\}\left(\frac{du(U)}{dU}\right)^{-1} = \{U(u),u\}\left(\frac{dU(u)}{du}\right)^{-1}, \ee
and substituting $u(U)$ from Eq.~(\ref{matching}). As long as $r_s \gg l$, then to leading order the result is the usual canonical case \cite{Hiscock}
\be F(u) = -\frac{\hbar}{24\pi}\left(\frac{2 r_s \left(2 U-r_s\right)}{\left(U-2 r_s\right){}^4}\right).\ee
Here $v_0 = 2r_s$.  The analytic full-order solution is in Appendix~\ref{sec:apxflux}, Eq.~(\ref{flux}). The thermal flux as the horizon is approached, $U\rightarrow 0$,
\be F_\textrm{Hawking} = \frac{\hbar}{192 \pi r_s^2}\,,\ee
suggests that equilibrium conditions expected of a long-lived (flux plateau, e.g.\ \cite{paper2}) radiating black hole will demonstrate a consequent Planckian distribution of particle count for some limited period of time.

\subsection{Evaporative Energy} 

Conformal symmetry provides the additional benefit of calculable total energy in the limit $l\ll r_s$, such that \cite{GLW}
\be E = \frac{\hbar c^3}{96\pi G M} \ln \frac{r_s}{l} = \frac{T_H}{12}\ln \frac{r_s}{l}.\ee
The exact total energy expression is in Appendix~\ref{sec:apxflux}, Eq.~(\ref{totalenergy}).  Since the energy is finite, the evaporation process eventually stops. 
Geometrically, the classical Schwarzschild vacuum dictates $l\ll r_s$, but here we can energetically motivate that small scale in order that the radiated energy is substantial enough to exhaust the black hole. Going further, using the full expression Eq.~(\ref{totalenergy}) one can see that as $r_s$ approaches $l$, the full mass will be radiated if  $l$ is  of  order $l_P$. The black hole mass is carried away by the radiation (along with the information via a turn-over of the Page curve, e.g.\  \cite{Hwang:2017yxp}). At least in the conformal case, this represents an improvement to the canonical case, where the energy emitted is infinite. To understand whether a remnant is left over, finite energy emission is insufficient; a constant red-shift in the modes in the asymptotic future will signal its presence. In our case there is no remnant, and the evaporation is complete.

\subsection{Particle Spectrum} \label{sec:spec} 

The beta coefficients are computed exactly in the s-wave sector with the effective potential ignored \cite{GLW}. The particle spectrum is found to be, up to a constant:
\be N_\omega = \frac{1}{e^{\omega/T}-1}\,,\ee
in leading order $r_s\gg l$, where $T = (4\pi r_s)^{-1}$, demonstrating a quasi-thermal radiation field.  The fact that the spectrum is not exactly thermal, due to non-leading order corrections, (but is analytically known) means the information is transported on the radiation.  

\begin{figure}[ht]
\centering
\includegraphics[width=3.2in]{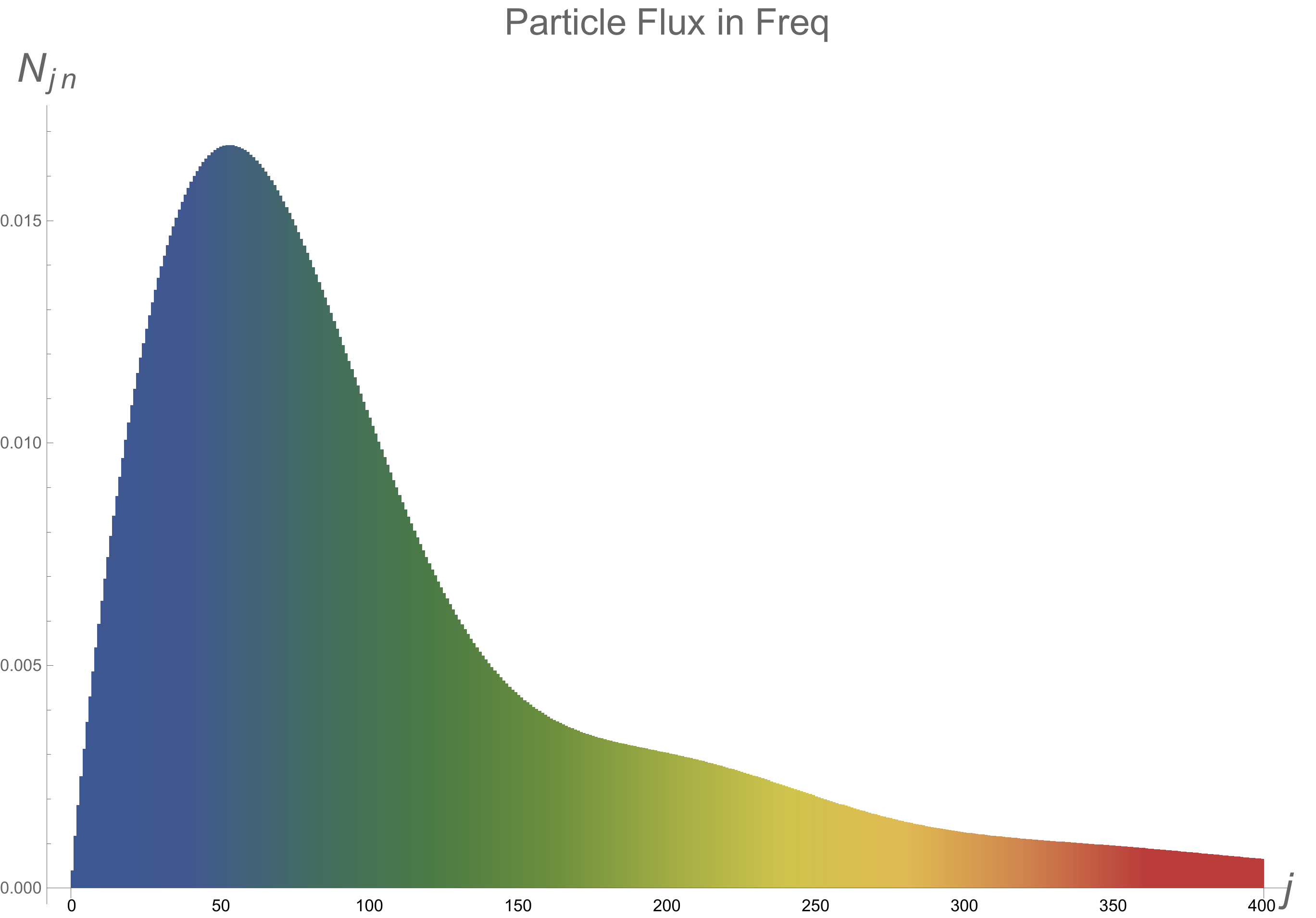} 
\caption{The discrete spectrum, $N_{jn}$, frequency evolved.  Here the system is set with surface gravity $\kappa = (4M)^{-1} = (2r_s)^{-1}=1$, and $2l=10^{-10}$.  The detector is set with $n=0$, $j=(0,400)$, $\epsilon = 0.001$.  The total particle count of this system is $N=2.209$ from numerically integrating over both `in' and `out' frequencies.  The sum of the displayed columns is $N=2.111$.  Inclusion of more frequency bins (`$j$') will yield the total count (e.g.\ \cite{signatures}).  
\label{fig:Particle_Flux_In_Freq}}
\end{figure}

The particular deviation from thermality also allows for a finite total amount of particle production, for a given $l$, which is readily calculated \cite{MPA} for a known black hole mass $M$.  
Frequency evolution can be resolved with the use of wave packets \cite{Hawking,Good:2013lca,Good:2016HUANG,wpU} confirming the total particle count done in \cite{GLW,MPA}.  With sufficiently small $l\ll r_s$, and good frequency resolution (small $\epsilon$) on the particle detector, one can see the spectrum shape well-resolved in Fig.~\ref{fig:Particle_Flux_In_Freq}.

\subsection{Temperature Correction via Surface Gravity} 

The temperature is not exactly the classical horizon 
temperature. We want the limiting temperature outside the black hole, $r>r_s$. 
Computing via the usual expression for surface gravity of a static metric, $2\kappa = g'_{tt}(r)|_{r\rightarrow r_s}$, and $\kappa = 2\pi T$, we have 
\be T = \frac{\Delta r_s}{4\pi \sqrt{\Delta^2 + l^2}\, (r_s + \sqrt{\Delta^2 + l^2})^2}\,.\ee
For $l = 0$, in the limit $r\to r_s$ (from above), the temperature is $T_H = (4\pi r_s)^{-1}$. In general the temperature is maximum right outside the Schwarzschild radius and can be found with an exact expression.  The corrected temperature to leading order is 
\be \frac{T}{T_H}= 1-3\left(\frac{l^2}{2r_s^2}\right)^{1/3}\,,\ee
cooler but generally a negligible decrease in temperature.

\subsection{Limitations of the Model}
It is worthwhile to comment on a few potential extensions of this work and current limitations of the proposed toy model.  In particular,
\begin{itemize}
    \item One expects a non-static situation, with increasing temperature associated with the decreasing black hole mass during evaporation.  This is not modeled.  However, it does not appear that this is intractable, e.g.\ via introduction of suitably creative matching conditions.
    \item There is non-trivial spatial curvature in the region just outside the black hole, $\Delta \equiv (r-r_s) \sim l$, i.e.\ within a few Planck lengths of the horizon. This is a regime beyond the scope of the current work. 
    \item Conformal asymmetry calculation for the stress tensor in $3+1$ dimensions and total energy production has not been attempted.  Besides issues of vacuum polarization, it is well known that conformal symmetry greatly simplifies the form of the stress tensor. 
    \item While the mode solutions are good approximations where $R\approx 0$, away from the horizon, they will not satisfy the corresponding wave equation if conformal coupling is chosen, $\xi = 1/6$, because $R\neq 0$ near the horizon. For minimal coupling, $\xi = 0$, the equation of motion is simple and the modes studied here are appropriate but the theory will not be conformally invariant in 3+1 dimensions in the massless limit.  
\end{itemize} 

Despite these extensions, the metric as is, Eq.~(\ref{QS}), is a conceptually transparent geometry for modeling some aspects of unitary black hole radiance.  The strength of this approach is the use of the preservation of unitarity (elaborated on in the next section) as a guiding principle with which to push the conventional model on the gravity side.  Rather than taking the matching condition as being dynamically determined by the matter, the matching condition is treated as a given.  While this may be considered a weakness, in the light of quantum pure evolution this approach appears as an additional strength accompanying the usual and simple quantum mechanical system of the moving mirror model. 

\subsection{Unitary Measure}
To characterize quantum purity one can take the key dynamical behavior of the mirror --  asymptotic inertia, which is responsible for information preservation -- and define a `topological rapidity', if you will, borrowing the notion that tearing is not allowed, only stretching.   This rapidity, $\eta$, can be used in the curved spacetime context to consider the asymptotic inertial geometric counterpart of the mirror's rapidity 
$\eta=\tanh^{-1}\beta$, where $\beta$ is 
the velocity in units of the speed of light. 

Helpful inspiration comes from the behavior of the quantum modes via the longitudinal relativistic Doppler factor:
\be D \equiv e^\eta = \sqrt{\frac{1+\beta}{1-\beta}}\,,\ee
which has a divergence if the emitting source is moving at the speed of light, $\beta=1$ (or zero if $\beta = -1$). In null coordinates, we see 
\be \frac{dv}{du} = \frac{d(t+x)}{d(t-x)} = \frac{1+\beta}{1-\beta} = e^{2\eta} = D^2\,. \label{eq:rapidc}\ee 

In our geometric context, between inside and outside coordinates, we can define in analogy  to Eq.~(\ref{eq:rapidc}), via regularity $U\leftrightarrow v$, a rapidity of the worldline of the origin, $u(U)$,
\be |\eta| = \frac{1}{2} \ln \frac{d u(U)}{dU}\,.\ee
Information loss can be thought of as a divergence in this rapidity for some time $U$.  Worldlines, after all, are considered time-like, $\eta \neq \infty$. 

In the black mirror-black hole case, 
\be |\eta| = \frac{1}{2} \ln \left(1-\frac{2r_s}{U}\right),\label{redline}\ee
one has a divergence for the time $U\rightarrow 0^-$.  The inside time ends at the shell.  Here the horizon is set at $v_H=0$, and $v_0 = 2r_s$. The modes are lost in between $U = [0,2r_s]$, corresponding to where the black hole singularity is located, absorbing the field: destroying it.  This is the same as saying all modes that fall in past $v_H=0$ get trapped in the black hole singularity.  This information loss is characterized by an imaginary value of rapidity (for cases with light-like travel see e.g.\ \cite{CW, spin, Hotta:2015yla}).

\begin{figure}[ht]
\centering 
\includegraphics[width=3.2in]{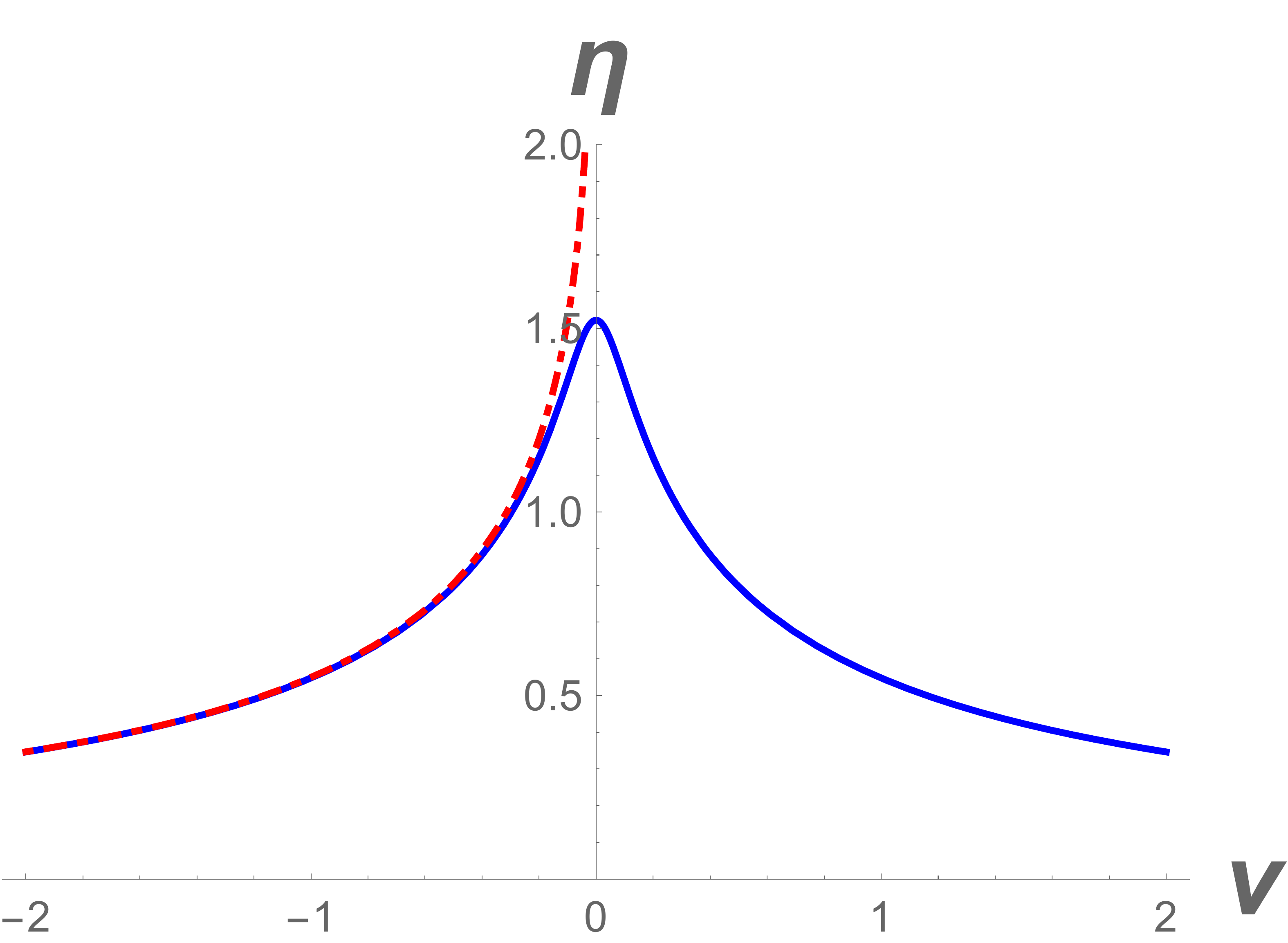} 
\caption{The dot-dashed red line, Eq.~(\ref{redline}), is the rapidity that characterizes information loss, in the classical case, for modes that fall past the horizon $v= v_H = 0$.  The horizon is set at $v_H = 0$, and the shell at $v_0=2r_s$.  Note the regularity condition $U\leftrightarrow v$. The solid blue line, Eq.~(\ref{blueline}), is the new rapidity $\bar{\eta}$ with no divergence, for $l = 0.05$. 
}
\label{fig:rapidity} 
\end{figure}

In the new geometry considered here, the rapidity of the origin, 
\be |\bar{\eta}| \equiv \frac{1}{2} \ln \left(1+\frac{2r_s}{\sqrt{U^2 + (2l)^2}}\right),\label{blueline}\ee
has no divergence for all real times $U$. In this chosen definition, we use the derivative taken with sign of $U$ as negative.  This sign will be fixed even when $U$ is positive, avoiding the divergence at $U = [0,2r_s]$; keeping the rapidity positive for $-\infty<U<+\infty$.  
The previous divergence at $U\rightarrow 0$ (or $v \rightarrow v_H =0$) is cured such that
\be |\bar{\eta}|_{U\rightarrow 0} = \frac{1}{2}\ln\left(1+ \frac{r_s}{l}\right)\,. \ee
A finite maximum topological rapidity, i.e.\ quantum purity in this context,
prevents the divergence.  The classical and new rapidities are plotted in Fig.~\ref{fig:rapidity}. 

It is worth noting that rapidity and entanglement entropy \cite{geoentropy} have a close relationship in the conformally symmetric context, $\eta = -6 S$, demonstrating that a divergence in entanglement entropy also signals information loss.

\section{Conclusion} \label{sec:concl} 

We extend the classical black hole with a toy model that preserves unitarity, demonstrating consistence with quantum mechanics, in 
quasi-thermal particle production from the semi-classical black hole background vacuum, to first order in the Planck length, i.e. corrections are higher order $\mathcal{O}(l^2)$. The black hole description tightens the analogy to the physics of an accelerated boundary (moving mirror).  This provides insights into how the equivalence principle applies from the emission of particles
in quantum theory \cite{pagefulling} to the geometric context of gravitation. 

The curved spacetime solution possesses several desired 
characteristics not present in previously known solutions, in particular giving 

\begin{itemize}

\item Construction of a geometry that reduces to the classical Schwarzschild metric but extends it with by incorporating another length scale, much smaller than the Schwarzschild radius, connected with the Planck length. This ``quantum'' Schwarzschild metric, Eq.~(\ref{QS}), is capable of describing the evolution of a pure state that remains unitary.
\item Demonstration that quasi-thermal equilibrium energy flux existing in the 1+1 conformally symmetric case hints that the 3+1 stress tensor may exhibit similar behavior consistent with the Planck distribution of the particle spectrum from the s-wave sector. \item Demonstration of finite energy in the 1+1 conformally symmetric case, implying that the radiation process stops, signaling an end to evaporation.  

\item Introduction of a modified matching condition, and a quantum Regge-Wheeler coordinate sufficiently describing field modes for a collapse to black hole model. The limits for the classical model are approached for $\mathscr{R} \equiv l/r_s \rightarrow 0$.  

\end{itemize}

Due to natural, quantum pure plane wave form modes in the early past and late future, it is clear no information is lost. We connect this to complete evaporation of the black hole, demonstrating the relevant coordinate transformations. The lack of soft particle 
divergence, the finite energy and particle count, and the lack of 
a remnant further point to this solution as an 
interesting and potentially fruitful laboratory for 
exploration of quantum particle production in the universe.

\acknowledgments 

Funding from state-targeted program ``Center of Excellence for Fundamental and Applied Physics" (BR05236454) by the Ministry of Education and Science of the Republic of Kazakhstan is acknowledged. MG is funded by the ORAU FY2018-SGP-1-STMM Faculty Development Competitive Research Grant No. 090118FD5350 at Nazarbayev University. 
EL is supported in part by the Energetic Cosmos Laboratory and by 
the U.S.\ Department of Energy, Office of Science, Office of High Energy 
Physics, under Award DE-SC-0007867 and contract no.\ DE-AC02-05CH11231.

\newpage

\onecolumngrid
\appendix
    
\section{Spacetime Quantities}\label{sec:apxcnx}

Here we present in detail many of the geometric objects of the metric Eq.~(\ref{QS}). We write $z \equiv +\sqrt{\Delta^2+l^2}$. 

\subsection*{Connections}
The Christoffel symbols are symmetric under interchange of the last two indices, so only the independent components are displayed. The indices 
1, 2, 3, 4 represent $t$, $r$, $\theta$, $\phi$. 
The results are 
\bea 
\Gamma^1_{21} &=& \frac{\Delta r_s}{2 z^2 \left(z+r_s\right)} \quad\rightarrow\quad \frac{r_s}{2r\Delta}\left [1+{\mathcal O}\left(\frac{l^2}{\Delta^2}\right)+{\mathcal O}\left(\frac{l^2}{r\Delta}\right)\right] \\ 
\Gamma^2_{11} &=& \frac{\Delta r_s}{2 \left(z+r_s\right){}^3} \quad\rightarrow\quad \frac{r_s\Delta}{2r^3} \left[1+{\mathcal O}\left(\frac{l^2}{r\Delta}\right)\right]\\ 
\Gamma^2_{22} &=& -\frac{\Delta r_s }{2 z^2 \left(z+r_s\right)} \quad\rightarrow\quad - \frac{r_s}{2r\Delta}\left[1+{\mathcal O}\left(\frac{l^2}{\Delta^2}\right)+{\mathcal O}\left(\frac{l^2}{r\Delta}\right)\right] \\ 
\Gamma^2_{33} &=& -\frac{zr}{z+r_s} \quad\rightarrow\quad -\Delta \left[1+{\mathcal O}\left(\frac{l^2}{\Delta^2}\right)+{\mathcal O}\left(\frac{l^2}{r\Delta}\right)\right] \\ 
\Gamma^2_{44} &=& -\frac{zr}{z+r_s}\sin ^2\theta \quad\rightarrow\quad -\Delta\sin^2\theta \left[1+{\mathcal O}\left(\frac{l^2}{\Delta^2}\right)+{\mathcal O}\left(\frac{l^2}{r\Delta}\right)\right]\ , 
\eea 
where the right arrow gives the lowest order correction in $l/\Delta$ and $l/r_s$, with the classical limit being  
$l=0$. 
The last four independent cases remain the usual classical Schwarzschild connections: 
\be 
\Gamma^3_{32} = \frac{1}{r}, \quad\quad \Gamma^3_{44} = -\cos\theta\sin\theta, \quad\quad  \Gamma^4_{42} = \frac{1}{r}, \quad\quad \Gamma^4_{43} = \cot\theta\ .\ee

\subsection*{Riemann Tensor} 
The nonzero components are displayed by the following expressions. Here we use $R^{\lambda}_{\mu\nu\sigma}$ format.  One can obtain, for example, $R^1_{231}$ from the $R^1_{213}$ using the antisymmetry of the Riemann tensor under exchange of the last two indices. The antisymmetry under exchange of the first two indices of $R_{\lambda \mu \nu \sigma}$ is not evident here because the components of $R^{\lambda}_{\mu\nu\sigma}$ are displayed.
\be 
R^{1}_{221} = \frac{r_s \left(-2 \Delta ^4+l^4-\Delta ^2 l^2+l^2 z r_s\right)}{2 z^5 \left(z+r_s\right){}^2} ,\quad\quad
 R^{1}_{331} = \frac{\Delta  r_s \left(\Delta +r_s\right)}{2 z \left(z+r_s\right){}^2} ,\quad\quad
R^{1}_{441} = \frac{\Delta   r_s \left(\Delta +r_s\right)}{2 z \left(z+r_s\right){}^2}\sin^2\theta \ee
\be
 R^{2}_{121} = \frac{r_s \left(-2 \Delta ^4+l^4-\Delta ^2 l^2+l^2 z r_s\right)}{2 z^3 \left(z+r_s\right){}^4} ,\quad\quad
 R^{2}_{332} = \frac{\Delta  r_s \left(\Delta +r_s\right)}{2 z \left(z+r_s\right){}^2} ,\quad\quad
R^{2}_{442} = \frac{\Delta   r_s \left(\Delta +r_s\right)}{2 z \left(z+r_s\right){}^2}\sin^2\theta \ee
\be
R^{3}_{131} = \frac{\Delta  r_s}{2 \left(\Delta +r_s\right) \left(z+r_s\right){}^3} ,\quad\quad
R^{3}_{232} = -\frac{\Delta  r_s}{2 z^2 \left(\Delta +r_s\right) \left(z+r_s\right)} ,\quad\quad
 R^{3}_{443} = -\frac{ r_s}{z+r_s}\sin^2\theta \ee
 \be
R^{4}_{141} = \frac{\Delta  r_s}{2 \left(\Delta +r_s\right) \left(z+r_s\right){}^3} ,\quad\quad
R^{4}_{242} = -\frac{\Delta  r_s}{2 z^2 \left(\Delta +r_s\right) \left(z+r_s\right)} ,\quad\quad
R^{4}_{343} = \frac{r_s}{z+r_s} \ee

\subsection*{Ricci Tensor} 

\be 
 R_{11} = \frac{r_s \left(l^2 \left(r_s-3 z\right)+\frac{2z^2  \left(\sqrt{z^2-l^2}+z\right) r_s}{\sqrt{z^2-l^2}+r_s}\right)}{2z^2  \left(r_s-z\right){}^4},\quad \quad
 R_{22} =\frac{-r_s \left(l^2 \left(r_s-3 z\right)+\frac{2 z^2 \left(\sqrt{z^2-l^2}+z\right) r_s}{\sqrt{z^2-l^2}+r_s}\right)}{2 z^4 \left(r_s-z\right){}^2}.
 \ee
 \be
 R_{33} = \frac{r_s \left(\left(\sqrt{z^2-l^2}+z\right) r_s-l^2\right)}{z \left(r_s-z\right){}^2}, \quad\quad
 R_{44} = \frac{ r_s \left(\left(\sqrt{z^2-l^2}+z\right) r_s-l^2\right)}{z \left(r_s-z\right){}^2}\sin ^2\theta.
 \ee

\subsection*{Einstein Tensor}
\be 
 G_{11} = \frac{r_s (z-\Delta ) \left(r+z\right)}{r^2 \left(r_s+z\right){}^3},
\quad\quad
 G_{22} = -\frac{r_s (z-\Delta ) \left(r+z\right)}{z^2 r^2 \left(r_s+z\right)}.
 \ee
 \be
 G_{33} = \frac{r_s (z-\Delta ) r \left(r_s \left(\Delta ^2+r_s (\Delta +z)+z^2+4 \Delta  z\right)+3 \Delta  z (\Delta +z)\right)}{2 z^3 \left(r_s+z\right){}^3}.
 \ee
 \be
  G_{44} = \frac{r_s (z-\Delta ) r \left(r_s \left(\Delta ^2+r_s (\Delta +z)+z^2+4 \Delta  z\right)+3 \Delta  z (\Delta +z)\right)}{2 z^3 \left(r_s+z\right){}^3}\sin ^2\theta .
\ee

\subsection*{Ricci Scalar} 

\be  R = \frac{r_s (z-\Delta ) \left(r_s \left(r_s \left(-2 \Delta ^2-r_s (\Delta +z)+z^2-5 \Delta  z\right)+(\Delta +z) \left(-\Delta ^2+4 z^2-6 \Delta  z\right)\right)+z (\Delta +z) \left(2 z^2-3 \Delta ^2\right)\right)}{z^3 r{}^2 \left(r_s+z\right){}^3}\ee 

At leading order in $l/\Delta$ and $l/r_s$, 
\be 
R=0-\frac{r_s\l^2}{r^5}\left(1+\frac{r_s}{\Delta}\right)^3\ . 
\ee 
Note that outside the black hole the Ricci scalar does not vanish, as in the classical case, reflecting the particle production. 
If the nonzero Ricci scalar is due to a particle flux, one might expect it to behave at large distances as $1/r^2$, but 
instead we have  $1/r^2\times  (r_{\rm sc}/r)^3$ where $r_{\rm sc}=\sqrt[3]{r_s l^2}$. This is vaguely reminiscent of braneworld gravity where a screening scale of the higher dimension effects occurs for  $r_{\rm sc}=\sqrt[3]{r_s r_c^2}$, where $r_c$ is the crossover scale related to the higher dimensional Planck length.  

Within experimental bounds, (e.g.\ Event Horizon Telescope or gravitational wave observations, see \cite{Giddings} and references therein for horizon scale modifications of the classical geometry of a black hole) the quantum Schwarzschild metric has the potential to be more consistent with quantum theory than its classical counterpart because unitarity is preserved in the gravitational collapse geometry.  By relaxing the strict definition of the classical $R=0$ vacuum of general relativity near the black hole horizon, information loss can be avoided.

\subsection*{Kretschmann Scalar}
\be K = \frac{r_s^2 \left(4 z^8 \left(r_s+z\right){}^4+4 \Delta ^2 z^6 r{}^2 \left(r_s+z\right){}^2+z^2 r{}^4 \left(r_s (z-\Delta ) (\Delta +z)+z^3-3 \Delta ^2 z\right){}^2\right)}{z^8 r{}^4 \left(r_s+z\right){}^6}\ee

\subsection*{Limits: $r\rightarrow \infty$, $r\rightarrow r_s$, or $l\rightarrow 0$}
\be \lim_{r\rightarrow \infty} R = 0, \quad\quad  \lim_{r\rightarrow r_s} R = 
\frac{2l^2+2lr_s-r_s^2}{lr_s(l+r_s)^2},\quad\quad \lim_{l\rightarrow 0} R = 0.\ee
\be \lim_{r\rightarrow \infty} K = 0,\quad\quad \lim_{r\rightarrow r_s} K = \frac{4 l^4+8 l^3 r_s+4 l^2 r_s^2+r_s^4}{l^2 r_s^2 \left(l+r_s\right){}^4},\quad\quad \lim_{l\rightarrow 0} K = \frac{12 r_s^2}{r^6}.\ee 

\subsection*{Leading Order at $r\rightarrow \infty$}
\be R = -\frac{l^2 r_s}{r^5}+\mathcal{O}(r^{-6})\,. \ee
\be K = \frac{12 r_s^2}{r^6} -\frac{40 l^2 r_s^2}{r^8}+\mathcal{O}(r^{-9})\,. \ee

\begin{figure}[ht]
\centering 
\includegraphics[width=3.2in]{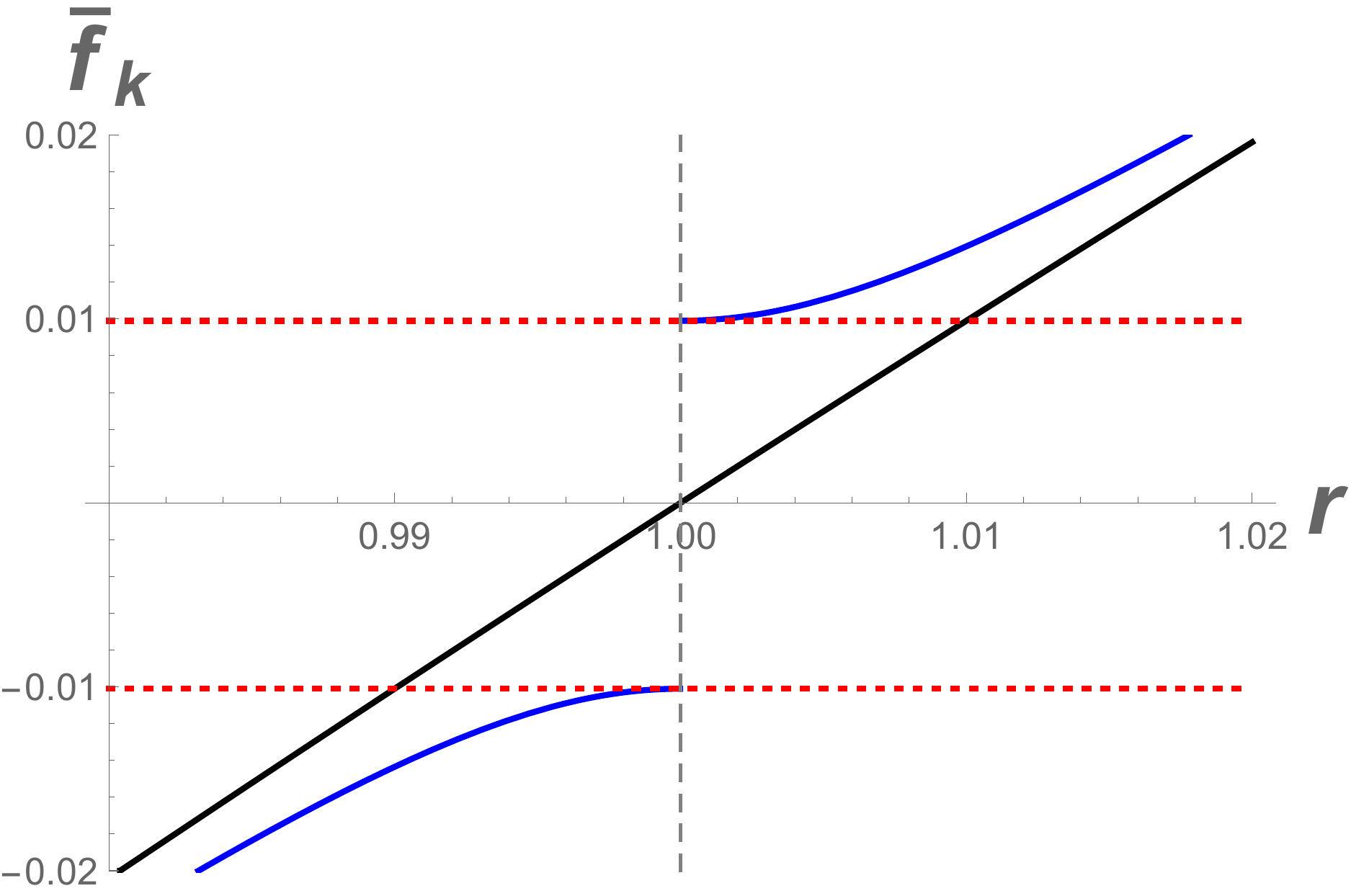} 
\caption{The black line is $f = 1-r_s/r$. The blue curves are $\bar{f}_k$.  Distances are in units of the horizon $r_s$, and results are shown for $l = 0.01$.  Horizontal dotted red lines are $ l/(l\pm r_s)$. }
\label{fig:fbar} 
\end{figure}

\subsection*{A Global Non-Collapse Geometry}

\be ds^2 = -\bar{f_k} dt^2 + \bar{f_k}^{-1}dr^2 + r^2 d\Omega\,,\label{ALLR}\ee 
Here
\be \frac{d\bar{r}^*}{dr}= \bar{f}^{-1}_k = 1+ k \frac{r_s}{z}\,,\ee
where $k= -1,0,+1$, for inside, on, and outside the Schwarzschild radius, i.e.\ $k = \textrm{sign}(\Delta)$. Figure~\ref{fig:fbar} plots  $\bar{f}_k(r)$. 
Note the removable discontinuity at the horizon.

\begin{figure}[ht]
\centering 
\includegraphics[width=3.2in]{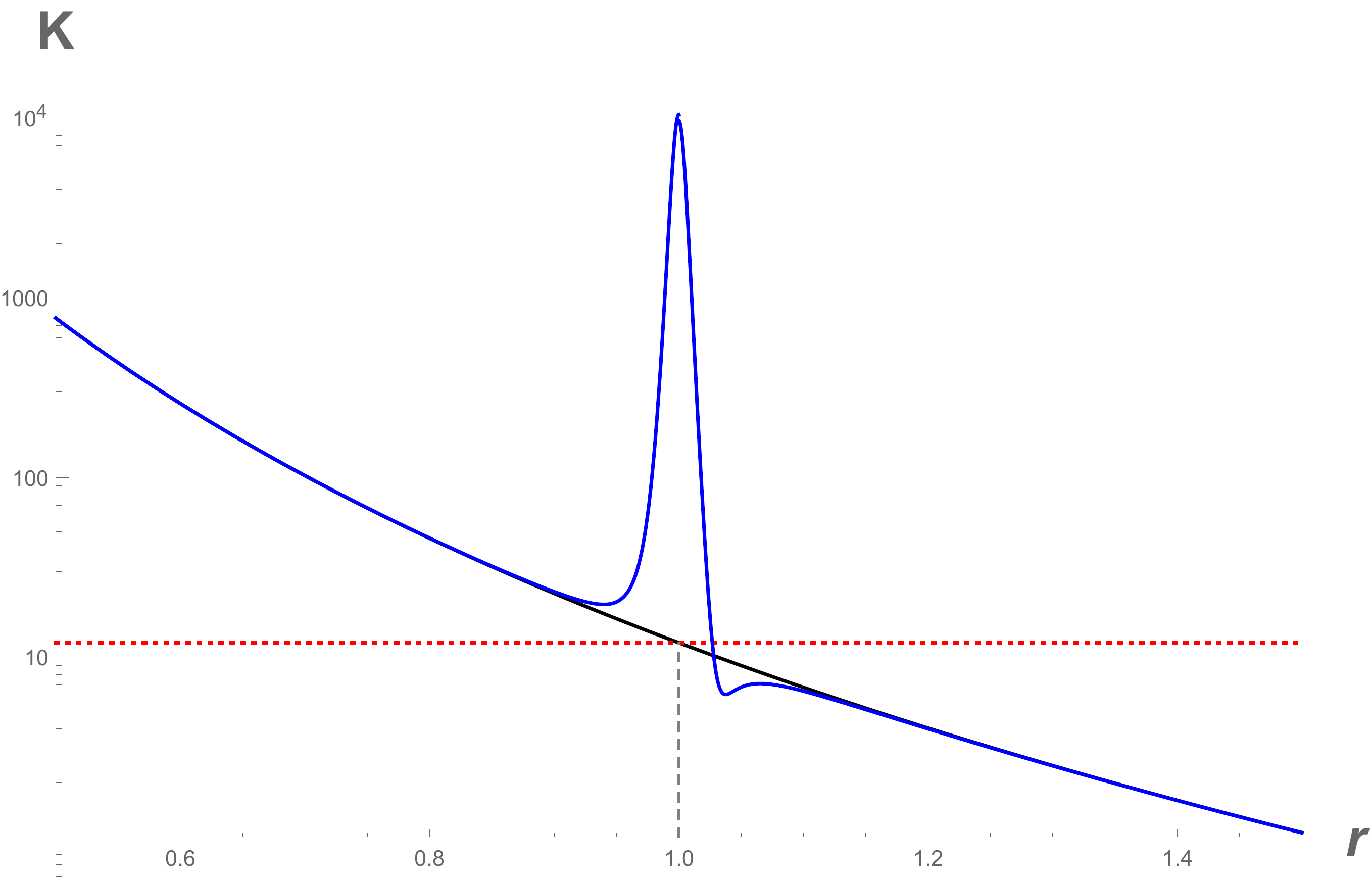} 
\caption{The black line is the Schwarzschild Kretschmann scalar $K =12r_s^2/r^6$, where $r_s = 1$.  The dotted red line is when it is at $r=1$, i.e. $K = 12$. The dashed vertical black line is at the horizon $r_s=1$. The blue curve is the quantum  $K$-scalar, with $l=0.01$.}
\label{fig:kvsr} 
\end{figure}

\begin{figure}[ht]
\centering 
\includegraphics[width=3.2in]{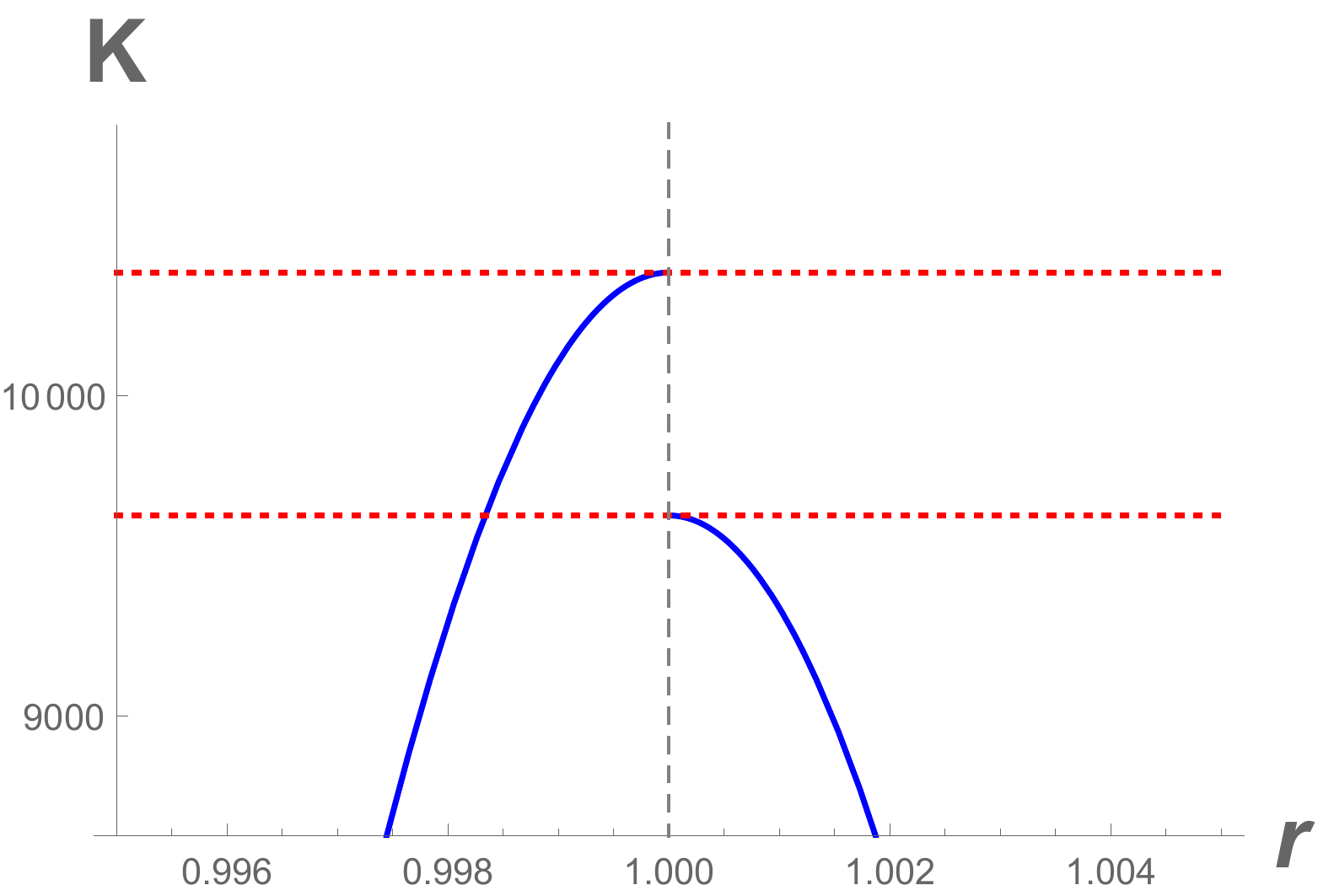} 
\caption{A zoom in of Fig.~\ref{fig:kvsr}. The dashed vertical black line is at the horizon $r_s=1$. The blue curves are the quantum $K$-scalar, near the removable horizon discontinuity.  The red dashed lines are at $\frac{4 l^4\pm 8 l^3 r_s+4 l^2 r_s^2+r_s^4}{l^2 r_s^2 \left(l\pm r_s\right){}^4}$, where $r_s=1$ and $l = 0.01$. }
\label{fig:kvsrzoom} 
\end{figure}

Figure~\ref{fig:kvsr} shows the Kretschmann scalar, in the quantum and classical versions. Again, there is modification near the horizon, and a small discontinuity at the horizon, proportional to $l$ -- best seen in the zoomed version Fig.~\ref{fig:kvsrzoom}.

\begin{figure}[ht]
\centering 
\includegraphics[width=3.2in]{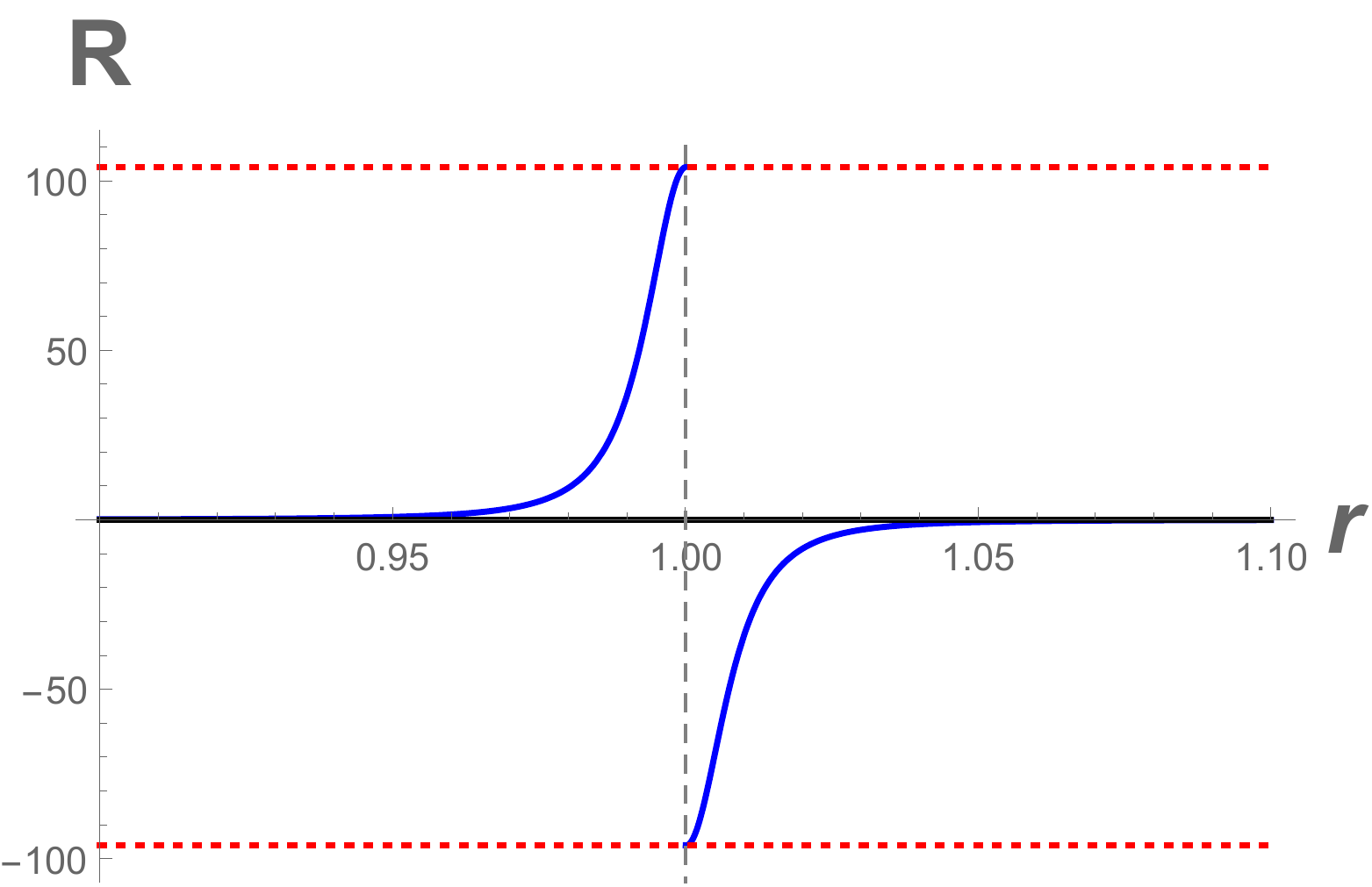} 
\caption{The dashed vertical gray line is at the horizon $r_s=1$. The blue curves are the quantum $R$-scalar, near the removable horizon discontinuity.  The red dashed lines are at $\frac{\pm 2 l^2+2 l r_s \mp r_s^2}{l r_s \left(l\pm r_s\right){}^2}$, where $r_s=1$ and $l = 0.01$. }
\label{fig:riccivsr} 
\end{figure}

The Ricci scalar is exhibited in Fig.~\ref{fig:riccivsr}, again showing modification near the horizon. Here the discontinuity is proportional to $1/l$.

\section{Inside Coordinate Collapse Geometry} \label{cover}


The canonical case of a collapsing shell of matter described both inside and outside will utilize the geometry of  Eq.~(\ref{QS}) and is expressed in null `inside' coordinates to give a complete cover of the spacetime, (see e.g. \cite{purity} for detail on outside coordinate representation), 
\begin{equation}\label{in-in}
\d s^2=
\begin{cases}
-\d U\d V,~~~~~~~~~~~~~~~~~~~~~~~~ \text{for}~v\leq v_0, \\
-\bar{f}(u,v)\bar{f}^{-1}(u,v_0)\d U\d V,~~\text{for}~v \geq v_0.
\end{cases}
\end{equation}
where $U$ and $V$ are the inside coordinates, as mentioned before,
\be U = T-r, \quad V = T+r, \ee
and the new tortoise coordinate, Eq.~(\ref{newtort}), helps us define $u$ and $v$ as the outside coordinates, as also previously given,
\be u = t-r^*, \quad v = t+r^*.\ee For clarity, and without loss of generality in 3+1 dimensions, we have dropped the $r^2 \d\Omega$ pieces of the metric on the 2-sphere.  The shell or shock wave is located at $v_0$, where the horizon $v_H \equiv v_0 - 2r_s$, is set to $v_H = 0$ for simplicity, so that $v_0 = 2r_s$. The metric is sewn on the shell $v_0$, from inside to outside:
\be \lim_{v_0 \rightarrow 2r_s} \bar{f}(u,v) \bar{f}^{-1}(u,v_0) = 1\,.\ee
The metric is also regular, like the canonical case, at the horizon $r=r_s$, but instead of $f=0$, we have $\bar{f} = l/(l+r_s) \approx l/r_s$.

\section{$1+1$, Energy Flux, Total Energy} \label{sec:apxflux} 

\subsection{From $1+1$ black mirror to $1+1$ black hole}
In the quantum pure black mirror \cite{GLW}, the calculations are most efficiently done using time as a function of space, $t(x)$, 
\be t(x) = -x - 2 l \sinh \frac{x}{r_s}\,, \ee
where we have anticipated the black hole collapse model by converting the parameters to the appropriate black hole quantities.  In $1+1$ dimensions, the single dimension of space, $x$, must be generalized to be more closely recognized in the radial black hole case, which can be done by inspection of its range, $+\infty$ to $-\infty$.  The matching condition, 
\be u(U) = U - 2 r_s \sinh^{-1} \left|\frac{U}{2l}\right|, \ee 
is re-arranged to give
\be |U| = 2l \sinh \left(\frac{U-u}{2r_s}\right), \ee
where we are now free to choose the negative sign because $U$ starts off negative at the beginning of collapse: 
\be U = -2 l \sinh \left(\frac{U-u}{2r_s}\right). \ee
Using regularity, $U\leftrightarrow v$, its clear that to describe the field modes our expression becomes in the outside null coordinates, 
\be v = -2l \sinh \left(\frac{v-u}{2r_s}\right), \ee
and in terms of space-time coordinates, we see
\be t(r^*) = -r^* - 2l \sinh \left(\frac{r^*}{r_s}\right). \label{eq:trstar}\ee
Therefore, in going from the $1+1$ dimensions of the mirror model to the $1+1$ dimensions of the spherically symmetric black hole collapse model, one replaces the spatial coordinate $x$ with the tortoise coordinate: $x\leftrightarrow r^*$.

\subsection{Energy Flux with NEF and Plateau; $1+1$ Case} 
Using the $t(r^*)$ function, and appropriately expressed coordinate transformed Schwarzian, the energy flux can be expressed as a function of $r^*$,
\be F(r^*) = \langle in|:T_{uu}:|in\rangle = \frac{r_s \left(4 l \cosh \left(\frac{r^*}{r_s}\right)-3 \text{sech}\left(\frac{r^*}{r_s}\right) \left[2 l+r_s \text{sech}\left(\frac{r^*}{r_s}\right)\right]+r_s\right)}{192 \pi  \left[l \cosh \left(\frac{r^*}{r_s}\right)+r_s\right]{}^4}\,.\label{flux}\ee
A plot of this function is given in Fig.~2 of \cite{GLW}, showing the thermal plateau and the negative energy flux spike.

\subsection{Finite Total Energy; $1+1$ Case}
The total energy is found by integrating the energy flux,
\be E = \int F du = \int_{+\infty}^{-\infty} F(r^*) \left(\frac{d t(r^*)}{dr^*} - 1\right) dr^*,\ee
over the Jacobian element, $du = dt-dr^*$, where the derivative is computed from Eq.~(\ref{eq:trstar}) as 
\be \frac{dt(r^*)}{dr^*} = -1 - \frac{2 l}{r_s}\cosh\left(\frac{r^*}{r_s}\right)\,.\ee
The result is 
\be E = -\frac{2 \left(6 l^4-9 l^2 r_s^2+2 r_s^4\right) \tanh ^{-1}\left(\frac{l-r_s}{\sqrt{r_s^2-l^2}}\right)+\sqrt{r_s^2-l^2} \left(3 \pi  l^3-6 l^2 r_s-3 \pi  l r_s^2+5 r_s^3\right)}{96 \pi  r_s^2 \left(r_s^2-l^2\right){}^{3/2}}\,.\label{totalenergy}\ee


\end{document}